\documentclass[twocolumn,showpacs,preprintnumbers,amsmath,amssymb]{revtex4}

\usepackage{graphicx}
\usepackage{dcolumn}
\usepackage{bm}
\newcommand{\be}{\begin{equation}}
\newcommand{\ee}{\end{equation}}

\begin{document}

\preprint{APS/123-QED}

\title{Effective boundary conditions for dense granular flows}

\author{Riccardo Artoni}
\email{riccardo.artoni@unipd.it}
\author{Paolo Canu}
\author{Andrea Santomaso}%
 
\affiliation{Dipartimento di Principi e Impianti di Ingegneria Chimica ``I. Sorgato''\\Universit\`{a} di Padova. Via Marzolo 9, 35100 Padova, Italy.
}%

\date{\today}

\begin{abstract}
We derive an effective boundary condition for granular flow taking into account the effect of the heterogeneity of the force network on sliding friction dynamics. This yields an intermediate boundary condition which lies in the limit between no-slip and Coulomb friction; two simple functions relating wall stress, velocity, and velocity variance are found from numerical simulations. Moreover, we show that this effective boundary condition corresponds to Navier slip condition when GDR MiDi's model is assumed to be valid, and that the slip length depends on the length scale that characterises the system, \emph{viz} the particle diameter. 
\end{abstract}
                              

\maketitle
Granular media exhibit a wide range of flow regimes \cite{jaeger96b}, as well as a plethora of dynamical instabilities \cite{king07}.
Focusing on gravity (or shear) driven flows, three regimes have been pointed out: (1) the collisional (\emph{gas-like}) regime, where energy is dissipated by the inelasticity of the collisions, (2) the dense flowing (\emph{liquid-like})regime, in which particles undergo long lasting contacts, and dissipation occurs through dynamic friction, and (3) the static (\emph{solid-like}) regime, which is capable to mantain structures due to the threshold, non-linear nature of static friction. These regimes were studied with both experiments and discrete models, the latter having experienced a great advance in the last years, starting from the work of Cundall and Strack\cite{cundall79}.\\Reliable continuum models would be of great advantage in simulating granular media, particularly when dealing with complex geometries or flows; in fact a unifying theory is still lacking. In this perspective, regimes (1) and (3) have been worked out with some success in a variety of theoretical studies, respectively with the kinetic theory of granular gases\cite{jenkins83} and with continuum critical state soil mechanics\cite{schofield}. For the dense regime, various theoretical approaches have been developed (and extensively reviewed in \cite{pouliquen02}); the last, more attractive one, is that proposed by the GDR MiDi based on the inertial number $I$ (see \cite{midi04}\cite{jop06b}\cite{pouliquen05}\cite{cruz05}), the importance of which was already stated by Goddard\cite{goddard86a}. However, despite the great attention towards continuum models and rheologies, little effort has been devoted to develop realistic boundary conditions for the velocity field at smooth or rough walls, even if the crucial role of side walls was recognized, for example, for inclined chute flows \cite{pouli05}. A common experimental approach developed to overcome this issue is the practice of gluing particles to the walls, in order to assume a no-slip boundary condition in the interpretation of the results. This intelligent choice is of fundamental importance but has, in our opinion, two major drawbacks: at first, it is known\cite{midi04} that for high shear rates particles undergo strong slip at the glued particles - bulk particles interface, a slip that adds some difficulty in holding the continuum hypothesis; thus it is not clear whether the glued particles are part of the bulk or of a bumpy wall, so that boundary conditions must be expressed on the first moving layer in contact with the glued one. The second drawback of this experimental practice is the partial applicability to real situations: the flow on smooth surfaces such as in hopper discharge usually shows particles slipping at the solid interface. Slip can be promoted or can be an undesired phenomenon, often we are concerned with stick-slip phenomena\cite{nasuno98}, which are common in dry-friction dynamics\cite{heslot94} ; in all of these cases, a deeper understanding of the behaviour of granular materials flowing near a boundary is needed, and the no-slip boundary condition is not the most valid approach.\\In a recent work\cite{artoni07} we used the mixing-length model proposed by the GDR MiDi and showed that using a slip boundary condition instead of a no-slip one considerably improved the predictions of the model in the vertical chute configuration; there we used a Coulomb friction condition, which could be a valid alternative to the no-slip condition. In this work we go even further, showing for a simple case that taking into account the effect of the heterogeneity of the force network yields an intermediate boundary condition which lies in the limit between no-slip and Coulomb friction; moreover, we show that this effective boundary condition corresponds to Navier slip-length condition if GDR MiDi's model is assumed, and that the slip length depends on the length scale that characterises the system, \emph{viz} the particle diameter.
We consider a single particle of mass $m$ and diameter $d$ lying on a plane, moving with instantaneous velocity $V$; the particle is subjected to a normal force $P$ and to a tangential force $T$. We will neglect, for simplicity, the effect of couples acting on the particle considering only traslational, sliding movements. We will assume that due to the heterogeneous nature of the medium the normal force $P$ is a random function of time with a given distribution function. Even $T$ could fluctuate, but we assume for the sake of simplicity that only the normal force does; qualitative results are not affected by the choice of the fluctuating force. Let $F$ be the friction force; we consider the simplest model of solid friction, e.g Amontons law, with only one friction coefficient $\mu$:
\begin{equation}
F=\left\{
\begin{array}{l}
T \; if\;\; V=0 \;and \; T<\mu P\\
\mu P  \;else\\
	\end{array}
	\right.
	\end{equation}
\begin{figure}[!h]
\centering{
	\includegraphics[height=3cm]{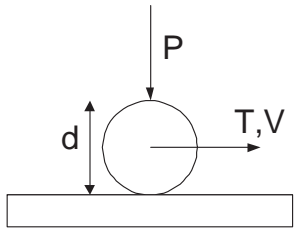}
	}
	\caption{Schematism of the variables considered in this work.}
	\label{velo}
\end{figure}
The motion of the particle is calculated from Newton's law:
\be
m\frac{d V}{d t}=T-F\label{newt}
\ee
If the normal force was constant, only two situations would be possible, corresponding respectively to no-slip and Coulomb conditions. But if the force fluctuated, the particle would undergo slip and no-slip events, which globally represent a non-Coulomb slip phenomenon; our aim is to derive an average expression for the slip velocity as a function of the forcings. Let's consider a typical distribution of normal forces of the form:
\be
p(f)=a (1-b e^{-f^2})e^{-\beta f}.
\ee
as suggested in \cite{mueth00}. This distribution of forces holds for normal forces in uniaxial compression, in a spatial sense; we will make the hypothesis that this distribution acts also between successive rearrangements of tangential forces in time. Our choice is supported by the fact that results do not depend on the particular choice of the distribution, apart from one point (the existence of a cutoff value in the force) which will be discussed later, and whose influence is limited. We suppose further that the force is a piecewise linear function whose nodes are extracted from this distribution. Let $P_{ave}$ be the average value of the normal force. We choose the time step between successive force rearrangements to be equal to the relaxation time $\tau= \sqrt{\frac{m\; d}{P_{ave}}}$\cite{midi04}; it follows directly that rescaling $t$ by $\tau$ the time step over which the force rearranges is $1$. Further rescaling leads to the dimensionless variables: $V'=V\sqrt{\frac{m}{P_{ave}\;d}}$ and $T'=T/\mu P_{ave}$, $P'=P/P_{ave}$. If we define $\alpha(t)$ as:
\begin{equation}
\alpha=\left\{
\begin{array}{l}
0 \; if\;\; V=0 \;and \; T<\mu P\\
1  \;else\\
	\end{array}
	\right.
	\end{equation}
the equation of motion becomes:
\be
\frac{d V'}{d t'}=\alpha\mu\left(T'-P' \right)\label{newt2}
\ee
from which we can compute the average rescaled slip velocity defined as:
\be
V'_{ave}=\mu \lim_{\tau\to+\infty} \frac{1}{\tau}\int_0^\tau{dt'\left[ \int_0^{t'}{ \alpha\left(T'-P'\right)dt}\right]}\label{vave}
\ee
We solve numerically the equation of motion; an example of the stick-slip behavior of the system is given in figure \ref{fig2}. An initially motionless particle can start to move only if the instantaneous normal force is below the yield threshold. A moving particle can decelerate only if the normal force is higher than the threshold. Moreover, it is clear from Fig. \ref{fig2} that the area in which normal forces oppose motion is larger than the area in which they promote motion; it is the dynamical nature of the system that causes the body to have a non null average velocity. It would be desirable to find a relationship between the average slip velocity computed by means of Eq. \ref{vave} and the rescaled average tangential force (which corresponds to a rescaled effective friction coefficient, being $\mu_{eff}/\mu=T/\mu P_{ave}$). After solving Eq. \ref{newt2} it is possible to look at the dependence of the statistics of the particle motion on the average value of the force in Fig. \ref{fig3}. 
\begin{figure}[!h]
\centering{
	\includegraphics[width=.9\columnwidth]{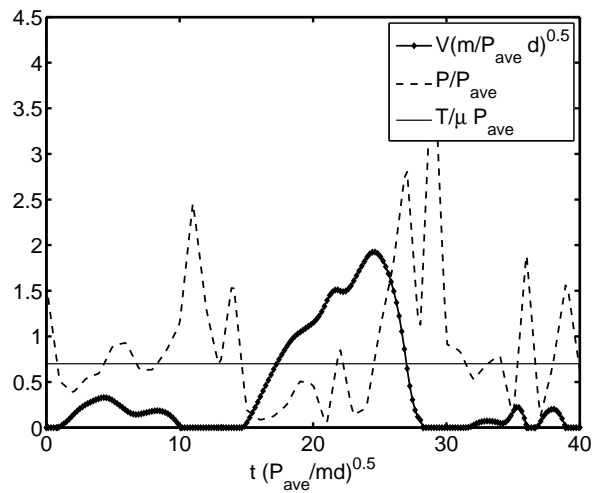}

	}
	\caption{Example of the local dynamics of the system.}
	\label{fig2}
\end{figure}
\begin{figure}[!h]
\centering{
	\includegraphics[width=.9\columnwidth]{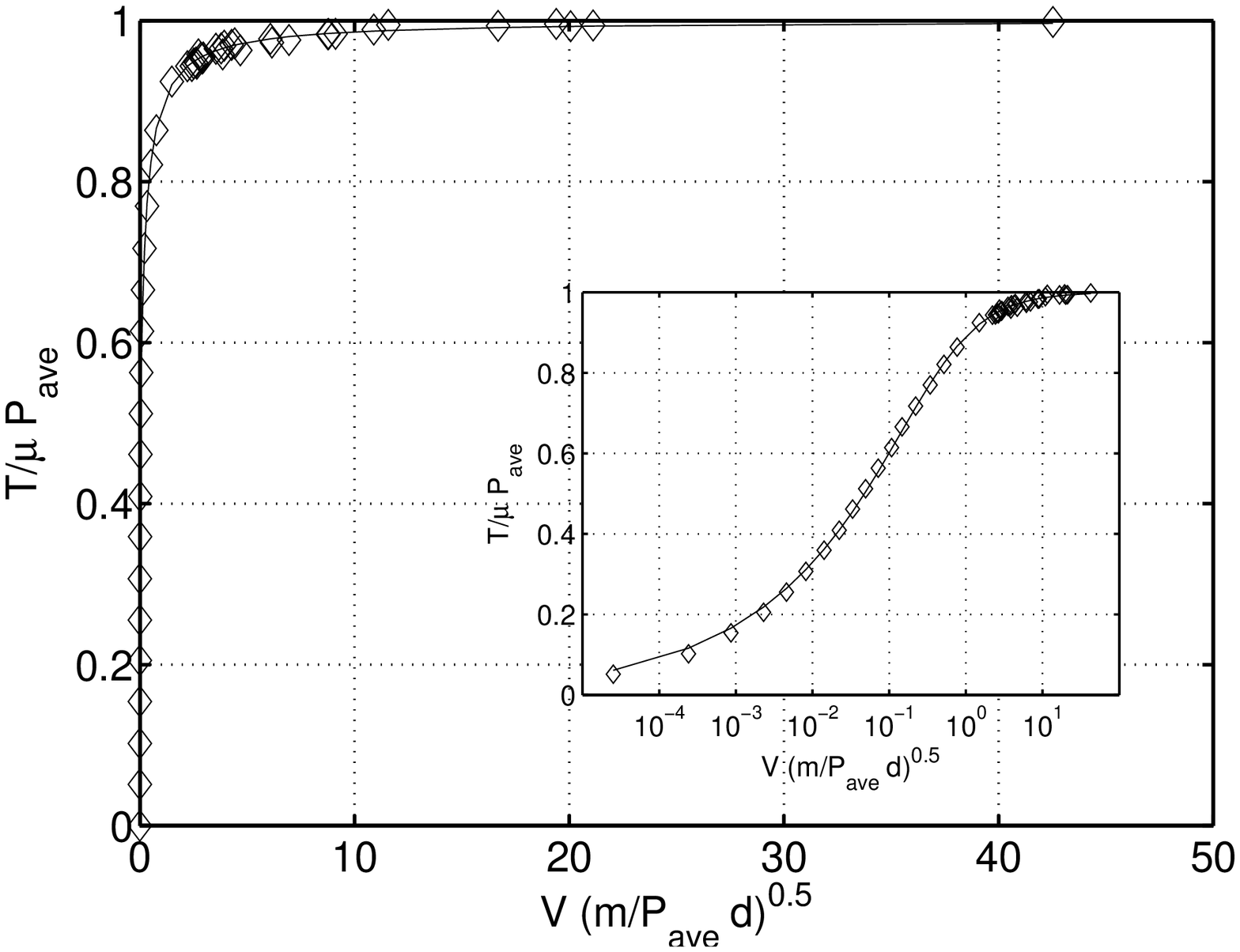}
	\includegraphics[width=.9\columnwidth]{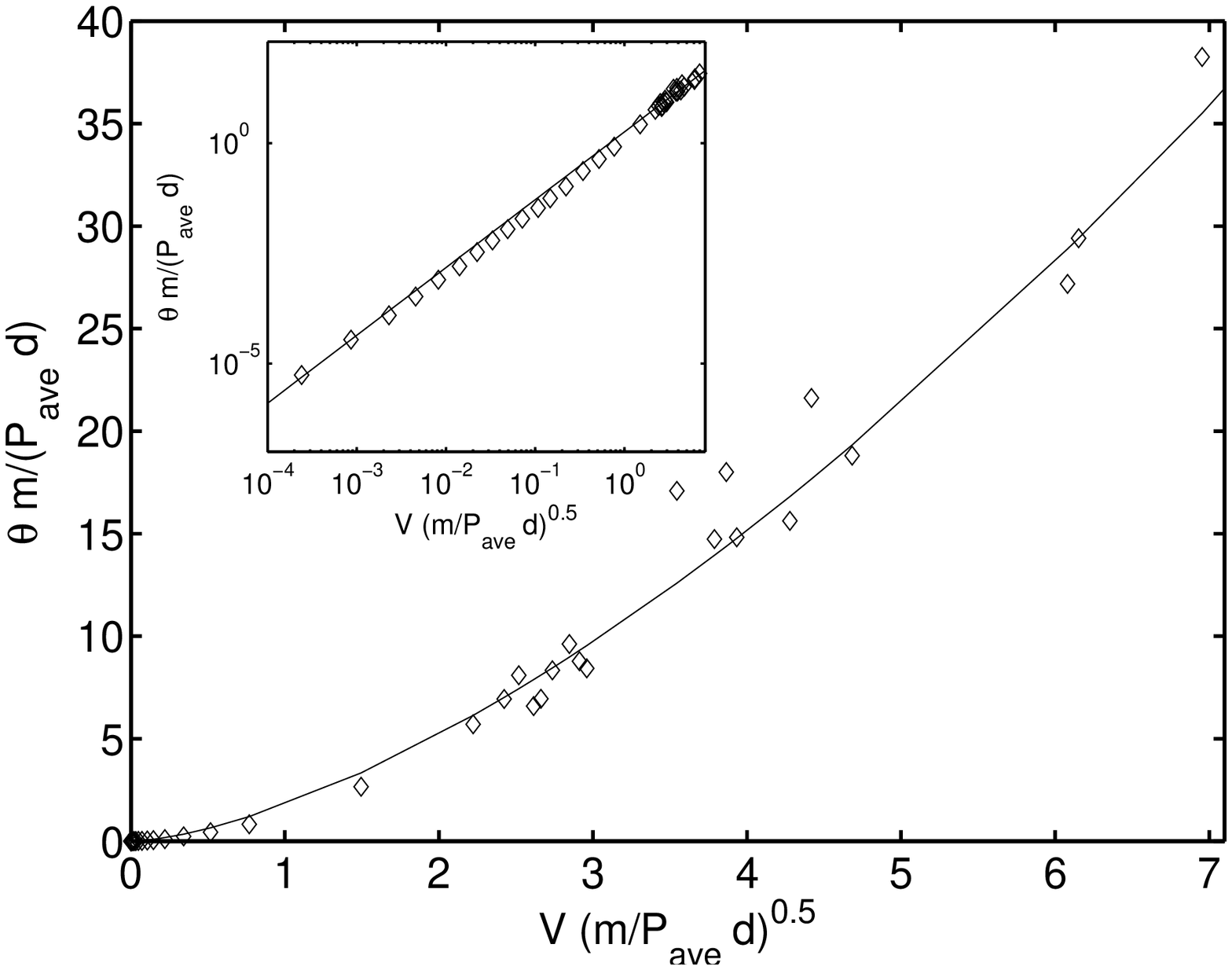}
		
	}
	\caption{Dependence of statistics of particle velocity on statistics of force. (top) Rescaled average slip velocity vs average pulling force. (bottom) Rescaled velocity variance vs  average slip velocity. Best fits from equations \ref{eq1},\ref{eq2} are also included.}
	\label{fig3}
\end{figure}
The curves evidence a no-slip limit at low values of the rescaled force $T'$, and a Coulomb limit for $T'\rightarrow 1$. The way $V'$ approaches $0$ depends on the nature of the distribution $p(T)$: if the distribution had an upper cut-off value, it would be easy to conclude that the system had a sort of yield-stress behavior at the wall, with a finite range of $T'$ giving $V'=0$; in the other case, without cut-off, the average velocity would be $0$ only for $T'=0$. This is the only point in which the choice of the distribution function qualitatively changes something in the results; however, the fast decrease of the tail in the distribution, if not giving a ``plastic'' behavior, would give some sort pseudo-plastic behavior, because of the need to impose a certain stress to obtain an appreciable slip. So, with a certain loss of exactness, it is possible to assume also a yield-stress formulation for the BC.\\It is interesting to note that also the variance of the distribution of the instantaneous particle velocities, that corresponds to the concept of granular temperature, which we express as $\theta=<(V(t)-V_{ave})^2>$, grows when $T'$ increases; so granular temperature at the wall has a behaviour similar to that of the slip velocity. Due to its definition, $\theta$ is made dimensionless with the position $\theta'=\theta \frac{m}{P\; d}$. In figure \ref{fig3} correlation between granular temperature and velocity is shown to follow a power-law behavior. From a general standpoint, the boundary conditions can be expressed with the help of the following fitting functions:\\
\be
T'=\left(\frac{V'}{V'+c_1}\right)^{\alpha}\label{eq1}
\ee
\be
\theta'=c_2 V'^\beta\label{eq2}
\ee
where $0<\alpha,\beta<0.5$ and $c_1,c_2>0$ are fitting parameters. A very good fit is obtained for $\alpha\approx 0.28$, $\beta\approx 1.5$, $c_1\approx 0.51$, $c_2=\approx 1.8$. The fit is the solid line in the figures. Eqq. \ref{eq1} and \ref{eq2} are the simplest expressions for the effective boundary conditions that can be applied at the wall characterised by a particle-wall friction coefficent $\mu$.\\
Navier boundary condition, relating the slip velocity and the gradient of the velocity normal to the boundary via a slip length $\lambda$, is a common way to characterise slip in fluid flows in micro and nanochannels; however, there is not a single plot of this condition in the $V'$ vs $T'$ diagram, because such a plot needs information on the relationship between stresses and deformation rates in form of constitutive relations. For a newtonian fluid, 
\be
V'=\lambda \frac{\mu}{\eta} \sqrt{\frac{m P}{d}}T'
\ee
which is linear and parametric in $ \frac{\mu}{\eta} \sqrt{\frac{m P}{d}}$. A Bingham yield-stress fluid will have an explicit relation of the form:
\be
V'=\lambda \frac{\mu}{\eta'} \sqrt{\frac{m P}{d}}\left(T'-T'_Y\right)\;\;\;(for\;\;\;T'>T'_Y)
\ee
where $T'_Y$ is the rescaled yield stress and $\eta'$ is the viscosity coefficient in Bingham's model. So a Navier condition for Bingham's model in the $V'$ vs $T'$ plot is a line shifted by $T'_Y$ and again parametric in  $ \frac{\mu}{\eta} \sqrt{\frac{m P}{d}}$. \\Both of these relationship obviously do not conform to the behavior obtained from the model developed in this work; assuming a mixing length model as GDR MiDi's, where $\frac{T}{P}=\mu(I)$, with $I=\frac{\dot{\gamma}}{\sqrt{P/m d}}$ (the difference in the expression of $I$ of the previous literature is due to the fact that here $P$ is a normal \emph{force}, not a pressure), the assumption of $\mu(I)=\mu_s+\frac{\mu_2-\mu_s}{I_0/I+1}$  (taken from Pouliquen and coworkers\cite{pouli05},\cite{jop06b})yields for a Navier BC:
\be
V'=\frac{\lambda}{d}\frac{T'-\mu_s/\mu}{\mu_2/\mu-T'} \;\;\;(for\;\;\;T'>\mu_s/\mu)\label{ciao}
\ee
which reaches an asymptote for $T'\rightarrow \mu_2/\mu$, and is $0$ in the range $0-\mu_s/\mu$. Thus, to unify the curves and represent the results obtained from the simple model of wall friction presented in this Letter, $\lambda$ must be a function of the form:
\be
\lambda=k\;d\; \zeta (T') \label{mlm}
\ee
where $\zeta(T')$ accounts for the change in the position of the asymptote and can be expressed simply as:
\be
\zeta=\frac{\mu_2/\mu-T'}{1-T'}
\ee
An important result is given in Eq. \ref{mlm}: to unify the curves as obtained in the ``experiments'', $\lambda$ must be a multiple of $d$: this is actually an important result, being $d$ the only internal length scale of the system, and so the best choice as a basis for estimating the slip length. The typical form of $V'\; vs\;T'$ curves for the various models is given in figure \ref{fig4}.\\
\begin{figure}[!t]
\centering{
	\includegraphics[width=.9\columnwidth]{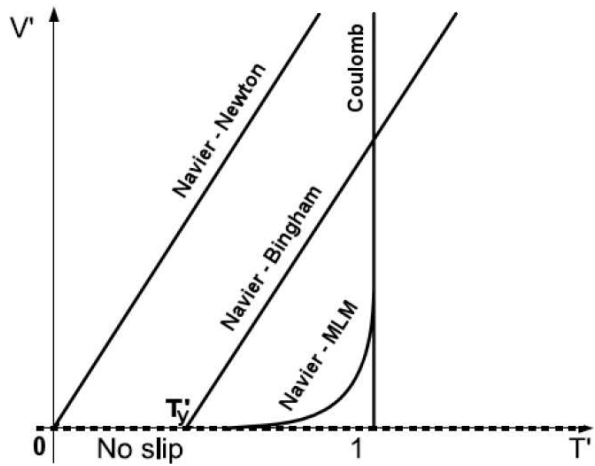}
	}
	\caption{Rescaled average velocity vs average pulling force for different BCs/constitutive laws. The slope of Newton and Bingham lines is $\lambda \frac{\mu}{\eta'} \sqrt{\frac{m P}{d}}$. }
	\label{fig4}
\end{figure}
To resume, the intermediate, efficient boundary condition we are looking for can be qualitatively expressed as Navier slip condition in a mixing length framework, the slip length corresponding to a multiple of the particle diameter. A step further can be made in the direction of determining a value for $\lambda$. Let's admit the yield-stress behavior of Pouliquen's form for $\mu(I)$, and suppose that $\mu_2\approx\mu$ (remember that $\mu$ is the particle-wall friction coefficient). In this perspective the slip length is simply proportional to the particle diameter and the best fit gives $\lambda/d\approx0.2$. This value gives a sort of minimum slip length; in the case with $\mu_2>\mu$ the slip length diverges for $T'\rightarrow1$.\\It is important to note that $V'$ is an analog of the inertial number $I$ for near wall flows, and $T'$ is an effective wall friction coefficient as $\mu(I)$ is for the bulk; thus it is interesting to note that  the shape of the curve $T'(V')$ is very close to that of $\mu(I)$; this can lead to some ideas on the origin of the effective friction coefficient in the bulk remembering that the effective wall friction coefficient derives from the assumption of heterogeneous forces.\\In this work we do not aim to define the correct functional form for these BCs (even if a very good fit was obtained for this simple case), but we want to underline that real boundary conditions (even in simplified setups) are not no-slip or Coulomb-like, and assuming one of these limiting BCs can introduce errors in the physical validity of granular flow models; this slip behavior can be captured by a modified Navier condition, where the slip length is proportional to the particle diameter.\\
To resume, a simple model of a particle sliding with the simplest frictional law on a plane has been developed in this Letter to determine effective boundary conditions for granular flows. To account for the heterogeneity of the medium, the particle is subjected to a random normal force, while a constant tangential force is assumed for simplicity. The dynamics consists of stick-slip events, which are related to the heterogeneity in the stress field; we reported the resulting dependence of the average tangential force on the average slip velocity and on the variance of the velocity of the particle (\emph{i.e.} granular temperature), thus providing two possible effective boundary conditions for the velocity and granular temperature fields. The results are well fitted by simple laws and represent for the velocity field an intermediate behavior between Coulomb's law (at high velocities) and the no-slip boundary condition. Granular temperature is related to the velocity by a simple power law behavior. In addition, we demonstrated that the curve obtained by numerical simulation satisfies a modified slip-length Navier boundary condition within a mixing length model of granular flow, with the slip length being proportional to the characteristic length of the system, the particle diameter.
 \bibliographystyle{nar}
 \bibliography{artoni_bc}
%


\end{document}